# Results of measurements of the flux of albedo muons with NEVOD-DECOR experimental complex

S.S. Khokhlov, N.S. Barbashina, A.G. Bogdanov, D.V. Chernov, V.A. Khomyakov, V.V. Kindin, R.P. Kokoulin, K.G. Kompaniets, A.A. Petrukhin, V.V. Shutenko, E.I. Yakovleva, I.I. Yashin
*National Research Nuclear University MEPhI (Moscow Engineering Physics Institute), Kashirskoe shosse 31, 115409 Moscow, Russia*

Results of investigations of the near-horizontal muons in the range of zenith angles of 85-95 degrees are presented. In this range, so-called 'albedo' muons (atmospheric muons scattered in the ground into the upper hemisphere) are detected. Albedo muons are one of the main sources of the background in neutrino experiments. Experimental data of two series of measurements conducted at the experimental complex NEVOD-DECOR with the duration of about 30 thousand hours 'live' time are analyzed. The results of measurements of the muon flux intensity are compared with simulation results using Monte-Carlo on the basis of two multiple Coulomb scattering models: model of point-like nuclei and model taking into account finite size of nuclei.

## 1. INTRODUCTION

Muon flux from the lower hemisphere is formed as a result of two processes: interactions of muon neutrinos passing through the Earth, and a scattering of atmospheric muons in the soil back into the upper hemisphere. Muons arising from the second process are called albedo muons. Such muons are the main source of the background in neutrino experiments. Theoretical estimates of the albedo muon flux depend on the models which give divergent results; so experimental data on atmospheric muons from the lower hemisphere are important for the physics of cosmic ray neutrinos.

## 2. ANALYSIS OF EXPERIMENTAL DATA

The experimental complex NEVOD (MEPhI, Moscow) is designed for investigations of all cosmic ray components on the Earth's surface, including albedo muons.

The basis of the complex is the Cherenkov water detector NEVOD [1] with a volume of $9\times9\times26$ m$^3$. The detecting system is formed by a spatial lattice of quasi-spherical modules (QSM), each of them including six PMTs with flat cathodes directed along the coordinate axes. The QSM lattice allows to detect Cherenkov radiation from any direction with practically the same efficiency. The lattice is formed by a set of vertical strings containing 3 or 4 modules each. The distances between the modules are 2.5 m along the detector, and 2.0 m both across it and over the depth. The current detecting system consists of 91 QSM (546 PMT in total).

To improve the event reconstruction accuracy, the coordinate detector DECOR [2] was constructed around the Cherenkov water calorimeter NEVOD (see figure 1). DECOR is a modular multi-layer system of plastic streamer tube chambers with resistive cathode coating. DECOR includes eight vertically suspended eight-layer assemblies (supermodules, SMs) of chambers with a total sensitive area of 70 m$^2$. The chamber planes are equipped with a two-coordinate external strip readout system that allows to localize charged particle tracks with about 1 cm accuracy in both coordinates ($X$, $Y$). The angular reconstruction accuracy for muon tracks crossing a single SM is better than 0.7° and 0.8° for projected zenith and azimuth angles respectively.

We have analyzed data of experimental series conducted at the experimental complex in 2002 – 2004 (10548 hours of 'live' time) and series conducted after modernization of the NEVOD measuring system [3,4] in 2011-2015 (19843 'live' time hours).

A condition of single near-horizontal muons selection was triggering of two DECOR supermodules located along the opposite short sides of the water tank. If track angles reconstructed on the basis of individual supermodule responses agreed within less than 5°, then it was supposed that the track segments within each SM belong to the same particle. The line connecting midpoints of track segments in both SMs is taken as the track of the muon. For such geometry of the experiment, muons with zenith angle in the range of 85°-95° are selected. The threshold energy of such muons is about 7 GeV.

Since the muon flux from the upper hemisphere is several orders of magnitude greater than the flux of muons moving upward (from below the horizon), a key element in detecting albedo muons is reliable determination of the direction of particles motion.

Two methods for determining the track direction can be used in the experimental complex NEVOD: measuring of Cherenkov radiation of charged particles in water and the time-of-flight technique.

Normally, most of hit photomultipliers are oriented toward the direction of muon arrival, so the direction of particle motion can be determined from the difference between the numbers of oppositely-oriented triggered photomultipliers. For selected events, it is convenient to use photomultipliers directed toward supermodules of the coordinate-tracking detector.

On the other hand, the triggering system of the coordinate-tracking detector contains information about the relative times of triggered supermodule actuation with a discreteness of 25 ns. The time of muon transit between SMs is about 90 ns, so the direction can be identified using the time-of-flight technique as well.

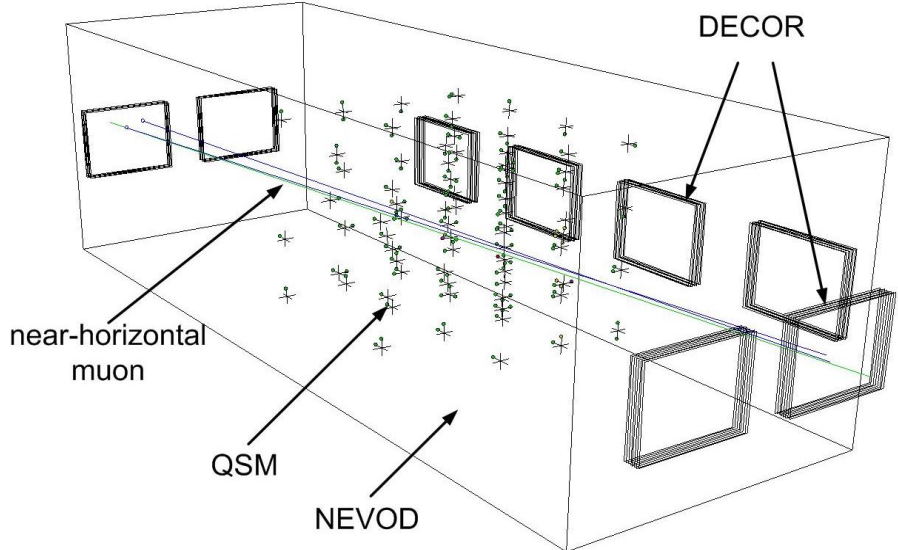

Figure 1: Experimental complex NEVOD.

An example of the event with albedo muon is presented on the figure 2. Inclined lines are tracks reconstructed on the basis of individual supermodules and the 'average' track. Crosses and circles represent triggered QSMs and PMTs. From the side of the right SM 35 photomultipliers were hit, from the opposite side only 5 PMTs; SM00 (right in figure 2) was triggered 150 ns before SM07 (left in the figure), so it is indicating that the muon came from below the horizon.

A cross-analysis of the two techniques revealed that the errors of amplitude technique are $\sim 10^{-3}$ for experimental series 2002-2004 and $\sim 10^{-5}$ for experimental series 2011-2015. The error of the time-of-flight technique is $\sim 10^{-3}$ for all series. When the amplitude and the time-of-flight techniques are combined, errors of determination of track direction are $\sim 10^{-6}$ and $\sim 10^{-8}$ for series of 2002-2004 and 2011-2015 respectively. The efficiency of event selection is ~96% [5].

## 3. INTENSITY OF THE MUON FLUX

By means of selected criteria, the direction for 5.46 million muon events was determined, among them 5717 albedo events with zenith angles of 91˚-95˚. For these events zenith and azimuth angles of the tracks were measured.

Distributions of the azimuth angle for muons coming from above (with zenith angle $\leq 89˚$), and from the bottom (with zenith angle $\geq 91˚$) are shown in figure 3.

Distinct central peaks in figure 3 correspond to the events in which muon passes through supermodules arranged strictly opposite to each other, right and left peaks correspond to cross-directed tracks. The distribution of albedo muons is not symmetric, which is determined by the topography of the surface in the surrounding area. Muons in the range of angles 75˚-105˚ come from the valley of the Moscow river. The deficit of muons in the range of angles 255˚-285˚ (most visible near 280˚) corresponds to muons coming from the side of a hill.

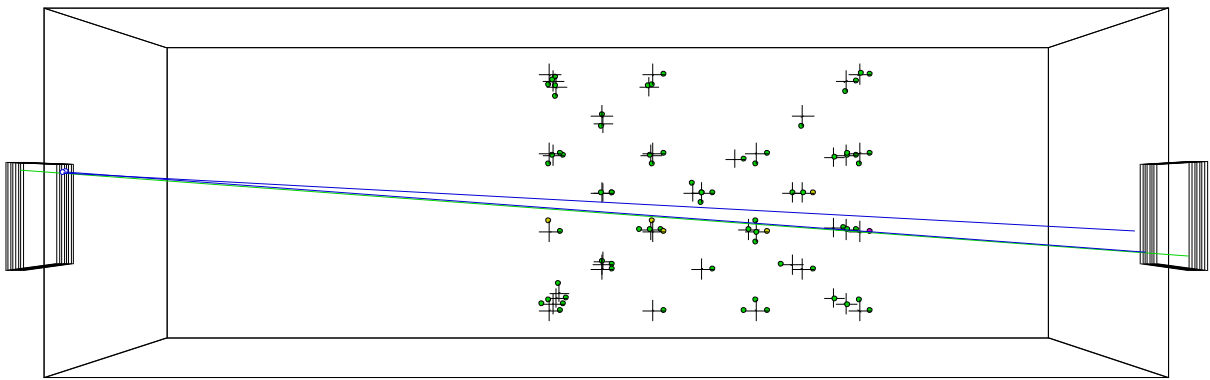

Figure 2: An example of the event with albedo muon ($\theta = 94.1°$).

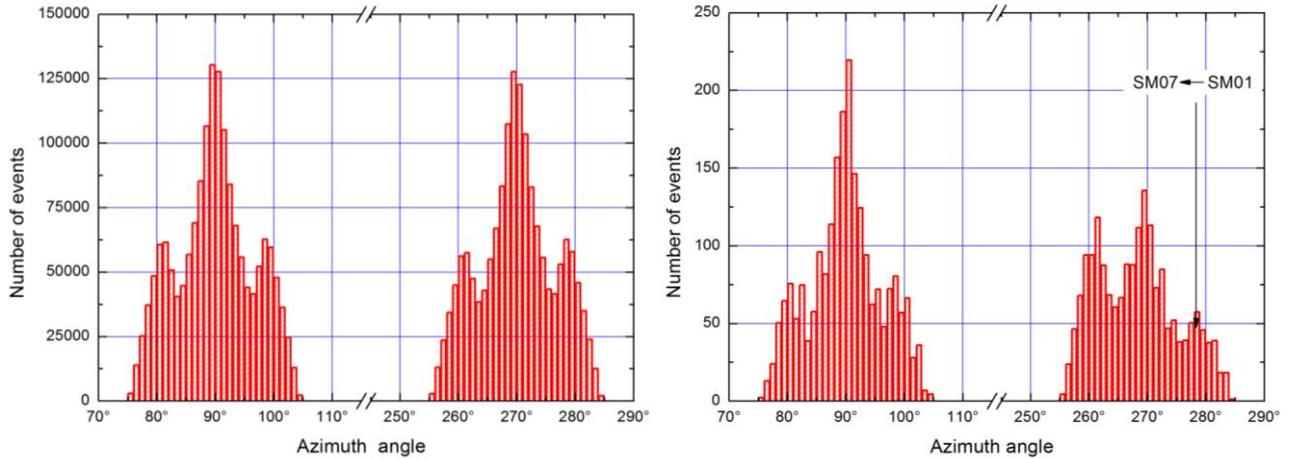

Figure 3: Azimuth angle distribution of near-horizontal muons for zenith angles ≤ 89˚ (left) and albedo events with zenith angles ≥ 91˚ (right).

The distributions of events over the measured range of zenith angles for two periods are presented in figure 4. The results of measuring series of 2002 – 2004 and 2011 – 2015 completely agree.

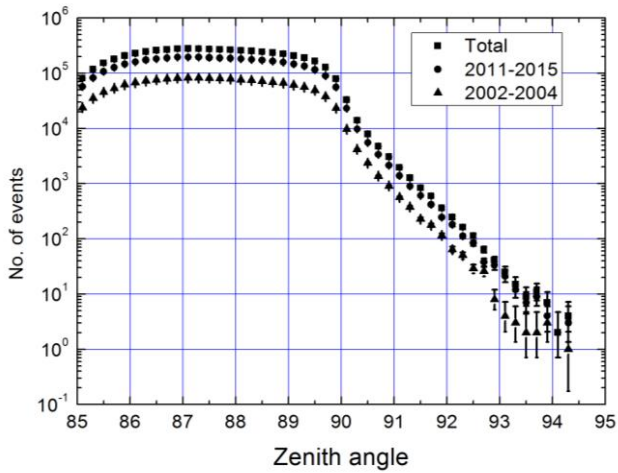

Figure 4: Zenith angle distribution of selected near-horizontal muons.

Taking into account the acceptance of DECOR, the zenith angle distribution was re-calculated into the intensity of muon flux near the horizon for two intervals of azimuth angles; the results are shown in figure 5.

Simulation of muon penetration through a flat layer of ground by means of Monte-Carlo method has been performed. In simulations, two models are used: point-like nuclei (Moliere theory [6,7]) and model taking into account finite size of nuclei according to [8]. Muon energy was simulated as a function of zenith angle based on differential spectrum of muons, calculated by the formulas from [9] and assumption about the power spectrum of primary cosmic rays with integral spectrum slope $\gamma = -1.7$.

Results of simulations are compared with experimental data in figure 5. The figure shows that the number of albedo muons for point-like nucleus model is in excess compared to the experimental results. The scattering model taking into account the finite size of nuclei is in a good agreement with the experimental muon flux from the side of relatively flat surface of the soil.

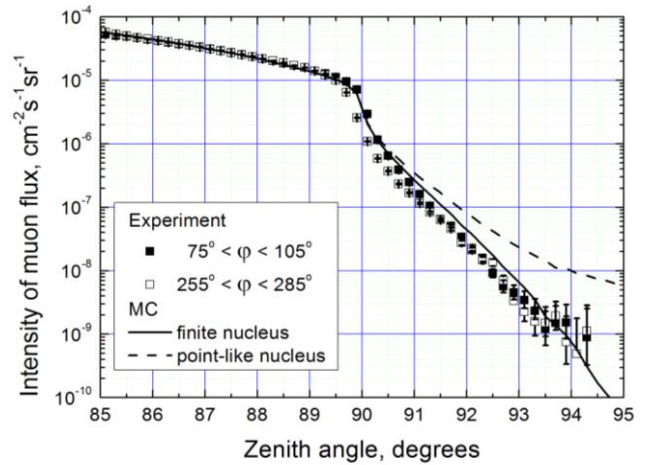

Figure 5: Intensity of muon flux.

## 4. CONCLUSION

As a result of experimental series with the total duration of ~30 thousand hours of 'live' time at the Experimental complex NEVOD, the intensity of atmospheric muons with energies >7 GeV in zenith angle range 85°-95° has been measured. It was found that the calculations for the scattering model of point-like nuclei (Moliere theory) lead to an overestimation of the calculated intensity of the albedo muons in comparison with experiment. The model taking into account finite size of nuclei is in a good agreement with the experimental data.

### Acknowledgments

The work has been performed at the Unique Scientific Facility NEVOD with the state support provided by the grant of the President of the Russian Federation (MK-7145.2015.2) and MEPhI Academic Excellence Project 02.a03.21.0005 of 27.08.2013.